\documentclass[12pt]{article}
\usepackage{amsmath}
\usepackage{amsfonts}
\usepackage{amssymb}
\usepackage{amsthm}
\usepackage{stackrel}

\usepackage{color}

\newcommand{\R}{{\rm I}\kern-0.18em{\rm R}}
\newcommand{\1}{{\rm 1}\kern-0.25em{\rm I}}
\newcommand{\E}{{\rm I}\kern-0.18em{\rm E}}
\newcommand{\p}{{\rm I}\kern-0.18em{\rm P}}
\makeatletter

\usepackage{epsfig}

\usepackage{fancyhdr}

\usepackage{graphics}

\usepackage{amssymb}
\usepackage{amsxtra}
\usepackage{amsfonts}

\usepackage{afterpage}
\usepackage{eucal}
\usepackage{eufrak}

\usepackage{enumerate}

\allowbreak

\author{Lev B Klebanov\footnote{Department of Probability and Statistics, MFF, Charles University, Prague-8, 18675, Czech Republic, e--mail: levbkl@gmail.com}}
\title{No Stable Distributions in Finance, please!}
\date{}

\begin{document}
\maketitle

\begin{abstract}
Failure of the main argument for the use of heavy tailed distribution in Finance is given. More precisely, one cannot observe so many outliers for Cauchy or for symmetric stable distributions as we have in reality. 

\noindent
{\bf keywords}:outliers; financial indexes; heavy tails; Cauchy distribution; stable distributions

\end{abstract}

\section{Introduction}\label{s1}
\setcounter{equation}{0}

Latest decades we observe a large massive of publications on the use of stable (and similar to them) distributions in Finance (see, for example, \cite{RM}, \cite{HB} and references there). 

{\it The purpose of this paper is to show that the use of heavy-tailed distributions in financial problems is theoretically baseless and the main argument used to support this use is just wrongly interpreted}.

As it was mentioned in \cite{KV}, the first argument usually arises when considering some stock index return data. The observed fact is that quite a lot of data not only fall outside the 99\% confidence interval on the mean (under normality of distribution assumption), but also outside the range of $\pm 5\; \sigma$ from the average, or even $\pm10\;\sigma$. On the basis of this observations two conclusions were made.

First (and absolutely correct) conclusion consists in the fact, that the observations under assumption of their independence and identical distribution are in contradiction with their normality. It is necessary to agree with this.

The second conclusion is that the distribution of these random variables is heavy-tailed. It was mentioned in \cite{KV} that this decision is not based on any mathematical justification. Furthermore, the fact mentioned above, is formulated incorrectly. Namely, one has to speak on empirical standard deviation $s$ instead of general standard deviation $\sigma$. We think, the authors understand this fact, because for heavy tailed distributions $\sigma = \infty$. However, it means, that ``outliers" here are understand in the following way: statistician observes many events of the form $\{|X_1|>k s\}$, where $s=(\sum_{j=1}^{n}X_j^2)/n-(\sum_{j=1}^{n}X_j/n)^2$ is a random variable, and $k$ is a constant. From my point of view, the heaviness of the tail of $X_1$ says nothing on probability of the event of interest: $\{|X_1|>k s\}$. Really, if random variable $X$ has heavy tails, it may take large values with higher probability than in Gaussian case. However, the values of $s$ are also higher, and we know nothing about the probability of the event $\{|X_1|>k s\}$. Below we shall try to say something about this probability to show, it is not high in the case of stable distributions, and for some values of $k$ just less than for normal distribution. If so, the first argument works against stable distributions too. However, the situation is not so easy. The behavior of the probability $\{|X_1|>k s\}$ depends on $n$, and the relations between this probabilities for Gaussian and other distributions may (and appears to be) different for different values on sample size $n$. 

One of the first scientists who proposed the use of stable distributions in Finance was Benoit Mandelbrot.
In his paper ``The Variation of Certain Speculative Prices" Benoit Mandelbrot wrote: ``Despite the fundamental importance of Bachelier's process, which has come to be called "Brownian motion," it is now obvious that it does not account for the abundant data accumulated since 1900 by empirical economists, simply because the empirical distribution so price changes are usually too "peaked" to be relative to samples from Gaussian populations. That is, the histograms of price changes are indeed unimodal and their central "bells" remind one of the "Gaussian ogive." The tails of the distributions of price changes are in fact so extraordinarily long that the sample second moments typically vary in an erratic fashion" (see \cite{BM}). However, it is not clear, how to define the notions ``outliers" and ``long tails". To define ``outliers", we have to make analysis of the probability of the event, mentioned above, namely, of the event $\{|X_1|>k s\}$, where the observations are supposed now to be non-normally distributed. Mandelbrot provided no calculations of this probability. The statement about ``long tail" is also informal. It is connected to central body of the distribution rather than on its tail. Below we shall see that the presence of outliers has no connection with tails character at all. 

\section{Main argument}\label{s2}
\setcounter{equation}{0}

Let us start with rather intuitive considerations. Namely, instead of the event  $\{|X_1|>k s\}$ let us consider corresponding  event for $t$-statistic: $\{|t_n|>k\}=\{ \sqrt{n}|\bar{x}|/s >k\}$ under assumption the the mean value of the observations is zero. Define 
\[ S_n= \sum_{k=1}^{n}X_k; \;\; V_n^2 =\sum_{k=1}^n X_k^2. \] 
It is known that Student $t$-statistic and a self-normalized sum $S_n/V_n$ have the same limit distribution \cite{Efr}. The asymptotic (as $n \to \infty$) distribution of this self-normalized sum was given by Logan, B. F., Mallows, C. L., Reeds, S. O. and Shepp, L. A. in \cite{LMRSh}. The expression of this distribution is very complicated. However, it has very simple expression of moments $\mu_{2k}$ in the case of symmetric $\alpha$-stable distribution with $1<\alpha \leq 2$. Namely, $\mu_2 = 1$, $\mu_4=1+\alpha$, $\mu_6=1+3\alpha+2\alpha^2$, $\mu_8 = (3+20 \alpha +34 \alpha^2+17 \alpha^3)/3$. We see that all given moments appears to be maximal for the case of normal distribution ($\alpha =2$). It gives us an idea that the probability of the event of interest in maximal for Gaussian case at least for limit distribution. 

In Table \ref{tab1} we provide numerical calculations of probability of the event  $\{|X_1|> k s\}$ for the case of three independent identical distributed random variables (so, $n=3$) for symmetric $\alpha$-stable distributions, ($\alpha$ changes from 1\footnote{Corresponds to Cauchy distribution} to 2\footnote{Corresponds to Gaussian distribution} with step 0.25; $k=3$ and $k=7$). 

\begin{table}[h]
\caption{ Probabilities of deviations larger than $3 s$}\label{tab1}
\begin{center}
\begin{tabular}{|l|l|l|l|l|}
\hline
$\alpha =1$ & $\alpha =1.25$ & $\alpha =1.5$ &$\alpha =1.75$& $\alpha =2$\\
\hline
 0.035214& 0.044251 & 0.0532881  & 0.060860 & 0.0690411\\
\hline
\end{tabular}
\end{center}
\end{table} 

In Table \ref{tab2} there are given the same probabilities as in Table \ref{tab1}, but for the case $k=7$.

\begin{table}[h]
\caption{ Probabilities of deviations larger than $7 s$}\label{tab2}
\begin{center}
\begin{tabular}{|l|l|l|l|l|}
\hline
$\alpha =1$ & $\alpha =1.25$ & $\alpha =1.5$ &$\alpha =1.75$& $\alpha =2$\\
\hline
 0.0045115& 0.0034141 & 0.0044689  & 0.0053509 & 0.010674 \\
\hline
\end{tabular}
\end{center}
\end{table} 

From both Tables \ref{tab1} and \ref{tab2} we see that for $n=3$ the probability of event of interest is the highest for the case of Gaussian distribution. However, it is not so for larger sample size. To understand the situation in general we used computer simulation. 

\begin{figure}[h]
\centering
\hfil
\includegraphics[scale=0.7]{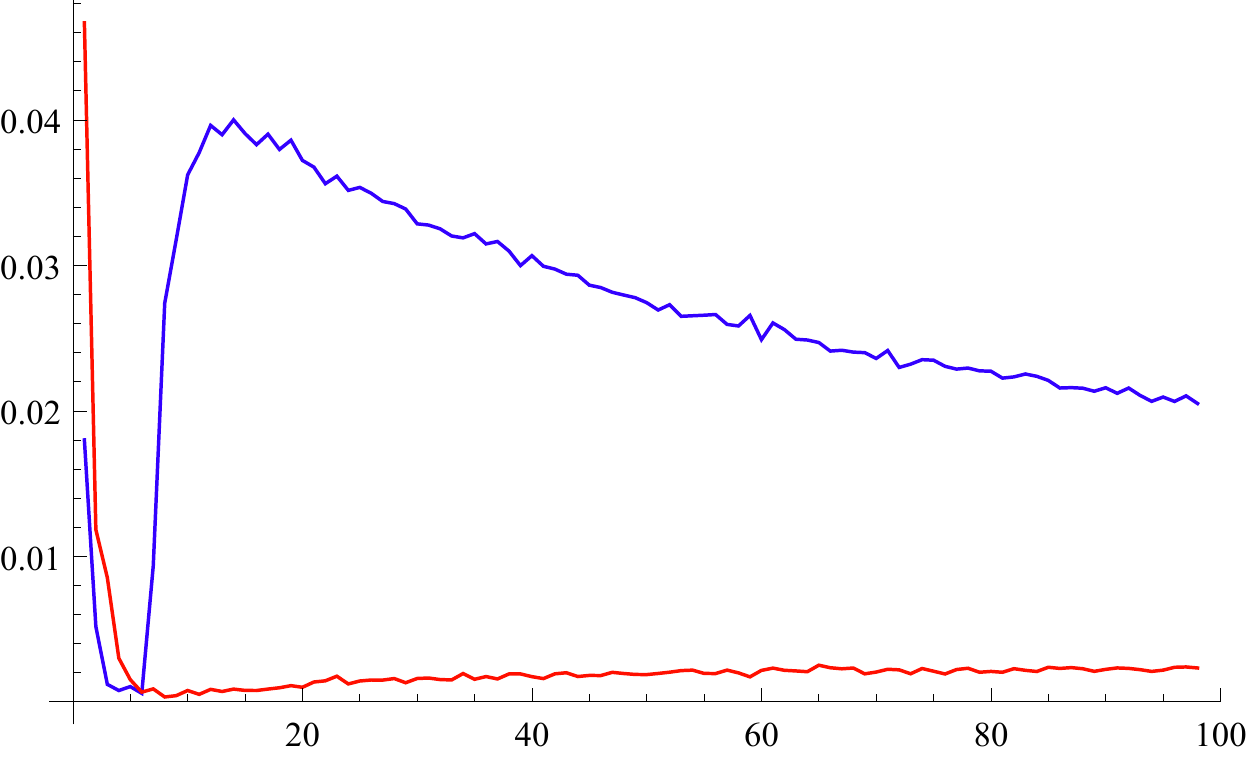}
\caption{Probability of event of interest depending on $n$ for $k=3$. Blue line corresponds to Cauchy distribution; red line - to Gaussian distribution}\label{fig1}
\end{figure}

For each $n$ there were simulated $M=1500$ samples of the volume $n$ each. For each sample we calculated the number of events $\{|X|>3 s\}$ divided by $n$. After that, this number has been averaged over all simulated samples and considered as an estimate of the probability of event of interest. The dependence of this probability on the sample size $n$ is given on Figure \ref{fig1}. The red line corresponds to Gaussian distribution, while the blue - to Cauchy distribution.

We see, that the blue line is below red one for $n=3;4$ only, while for the other given values of $n$ it is above red line. However, the probability of the event $\{|X|>3 s\}$ for Cauchy distribution decreases with increasing $n$. The sample size $n=100$ is rather small for financial indexes. Therefore, it is interesting to estimate the probability $\{|X|>3 s\}$ for much larger values of $n$. 

\begin{figure}[h]
\centering
\hfil
\includegraphics[scale=0.7]{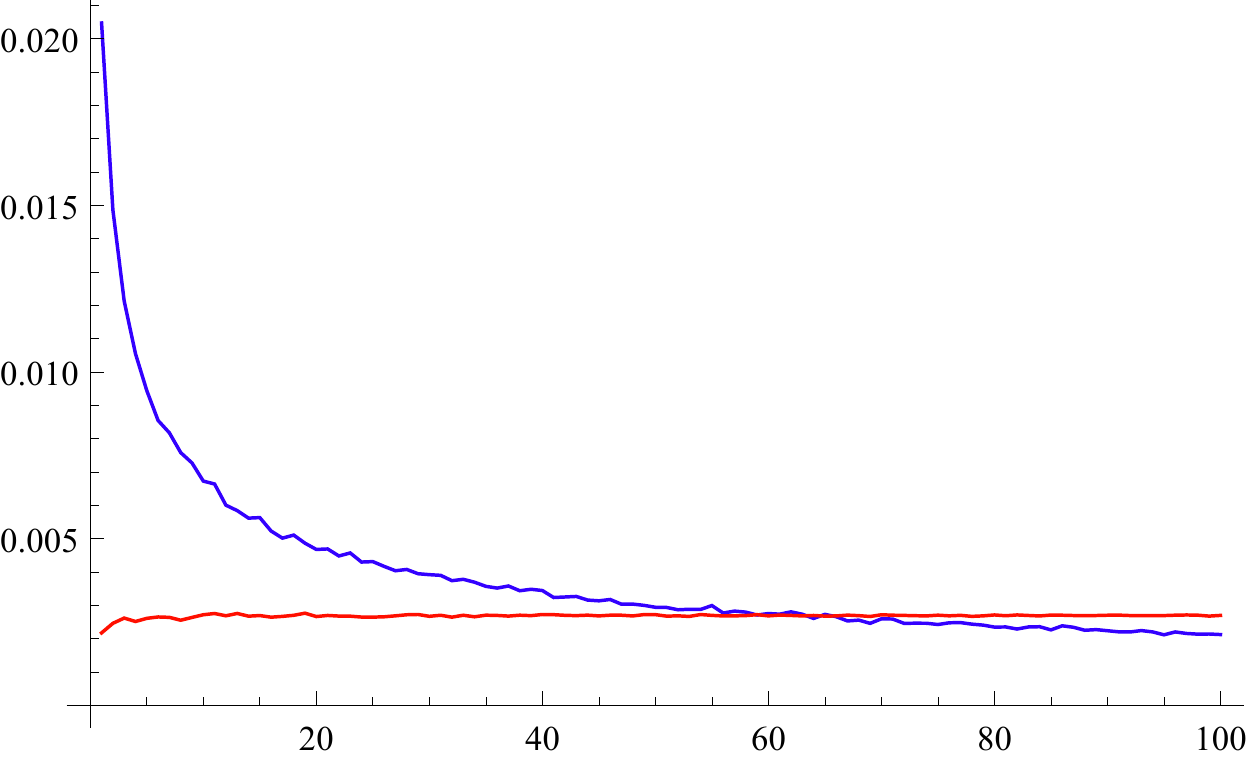}
\caption{Probability of event of interest depending on $n$ for $k=3$. Blue line corresponds to Cauchy distribution; red line - to Gaussian distribution. Sample size changes from 100 to 10000 with step 100.}\label{fig2}
\end{figure} 

\begin{figure}[h]
\centering
\hfil
\includegraphics[scale=0.7]{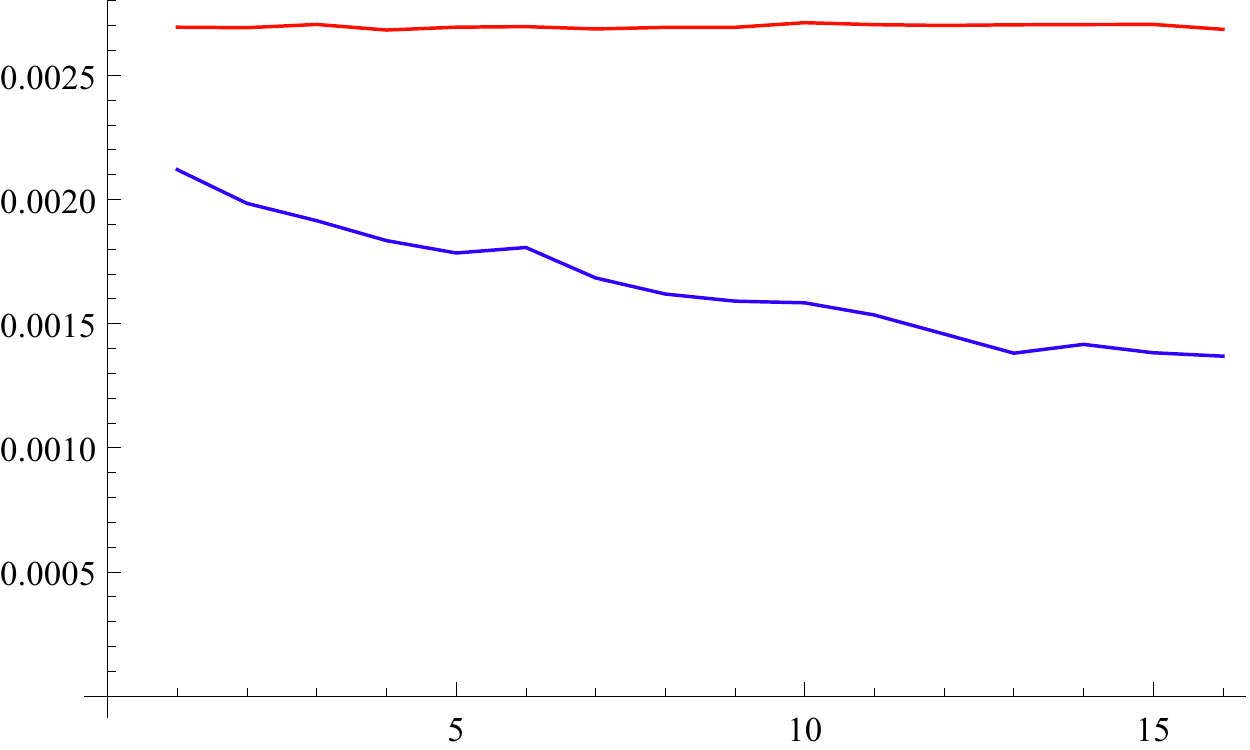}
\caption{Probability of event of interest depending on $n$ for $k=3$. Blue line corresponds to Cauchy distribution; red line - to Gaussian distribution. Sample size changes from 10000 to 25000 with step 1000.}\label{fig3}
\end{figure}

On the Figure \ref{fig2} there are given the dependences of the probability $\{|X|>3 s\}$ on $n$ for Cauchy (blue line) and Gaussian (red line) distributions. Sample size $n$ changes from 100 to 10000 with step 100. We see that for $n=6000$ the curves are very close to each other, and for $n>7500$ the probability of  $\{|X|>3 s\}$ is smaller for Cauchy distribution that for Gaussian. To see, that the probability remains to be smaller for Cauchy distribution that for Gaussian we provide more simulations with the results given on Figure \ref{fig3}. On Figure \ref{fig3} blue line corresponds to the probability of the event of interest for Cauchy distribution; red line corresponds to the probability of this event in the case of Gaussian distribution. Sample size changes from 10000 to 25000 with step 1000. We see, that the blue line is essentially below red one. The sample size $n=25000$ is not too small for financial indexes. However, we made calculations for sample size changing from 25000 to 60000. The result remains the same: probabilities of the event of interest remain to be smaller for Cauchy distribution than for Gaussian one (we omit the plots because they are similar to that in Figure \ref{fig3}).

\newpage

{\it This teaches us two lessons. The first is that the use of Cauchy distribution instead of Gaussian is senseless according to the reason advanced against Gaussian distribution itself. The second consists in the fact that probability of interest has no connection with heaviness of the tails of the distribution}. 

\section{Other stable distributions}\label{s3}
\setcounter{equation}{0}

Although Cauchy distribution is one representative (with $\alpha =1$) of the class of stable distributions, one may hope the situation with other stable distributions is different. Let us mention some results of simulations for other symmetric stable distributions. Looking on all above, we have some reasons expect similar behavior of the probability of interest in the of symmetric $\alpha$-stable distributions with arbitrary $\alpha \in (1,2)$. However, the value of sample size $n$ for which this probability for stable distribution will be close that for Gaussian case is expected to be dependent on $\alpha$. Simulations support this expectation. 

For $n=15000$ in the case of symmetric stable distribution with parameter $\alpha=1.2$ we have the probability of event of interest $|X_1|>3 s$ equal to 0.002862 versus  0.002697 for Gaussian distribution. 
We see, the difference is not high. For the case of $n=25000$ this probability is 0.002243 for $\alpha=1.2$ symmetric stable distribution versus 0.002697 for Gaussian case. Now the probabilities remains close to each other, but smaller for stable distribution than for Gaussian. The values of sample size for stable $\alpha=1.2$ distribution to have the probability of interest close to that of Gaussian distribution is essential larger than that for the case of pair Cauchy and Gaussian distributions. It seems to be natural, that this sample size value will increase with $\alpha$. This is so in reality. Therefore, we may expect that for some values of $\alpha$ close to 2 the value of $n$ giving identical values of probability of interest for $\alpha$-stable distribution and Gaussian one will be enormously large. In this situation we cannot say, that this stable distribution provides less outliers that Gaussian law. But will be the corresponding number of outliers sufficient to be in agreement with observed data? Unfortunately, there are only a few papers giving observed number of outliers, that is an estimator of probability of interest $|X_1|>3 s$. Some data of this kind may be found in \cite{EK}, Table 3. Corresponding estimators for probabilities $|X_1|>3 s$ given there is varying from 0.009 to 0.013. For symmetric stable distribution with $\alpha =1.8$ this probability changes from 0.006729 for $n=15000$ to 0.006356 for $n=25000$. We see that this probability is to small to explain the number of outliers in Table 3 from \cite{EK}.

One of the arguments against previous results given above is that we estimated the probability of the event
$|X_1|>3 s$, while we need to estimate $\p \{|X_1 -\bar{X}|>3s \}$. However, for the case of stable distributions with $\alpha \in (1,2)$ the law of large numbers is still true. Therefore, $\bar{X}$ is close to the mean value of $X$, and is small for symmetric case. We can support this by simulations. Namely, for the case of symmetric stable distribution with $\alpha =1.2$ and the number of observations $n=25000$ the probability  $\p \{|X_1 -\bar{X}|>3s \} \approx 0.002327$ what is close to that obtained without taking into account $\bar{X}$. This probability remains decreasing with growth on $n$. For example, for $n=35000$ it is  $\p \{|X_1 -\bar{X}|>3s \} \approx 0.002065$.

Another counter argument is that we have non-symmetric stable distributions in real problems. However, we  may consider symmetric case only. Really, if random variable $X$ has a stable distribution, then the difference $Z=X-Y$ has symmetric stable distribution for the case when $Y$ is independent copy of $X$. For real data we can obtain a sample from distribution of $Z=X-Y$ by random splitting of the sample into two equal parts and considering the differences between corresponding observations. We see, the number of outliers changes only a little. But now we are in symmetric situation, and may consider for outliers the values $Z$ for which $|Z|>3s$. 

Finally in this section we conclude that stable model cannot explain the number of outliers observed in real data from \cite{EK}. For some values of $\alpha$ this model has to be rejected according the same reasons as Gaussian model.

\section{What to do?}\label{s4}
\setcounter{equation}{0}

\begin{figure}[h]
\centering
\hfil
\includegraphics[scale=0.7]{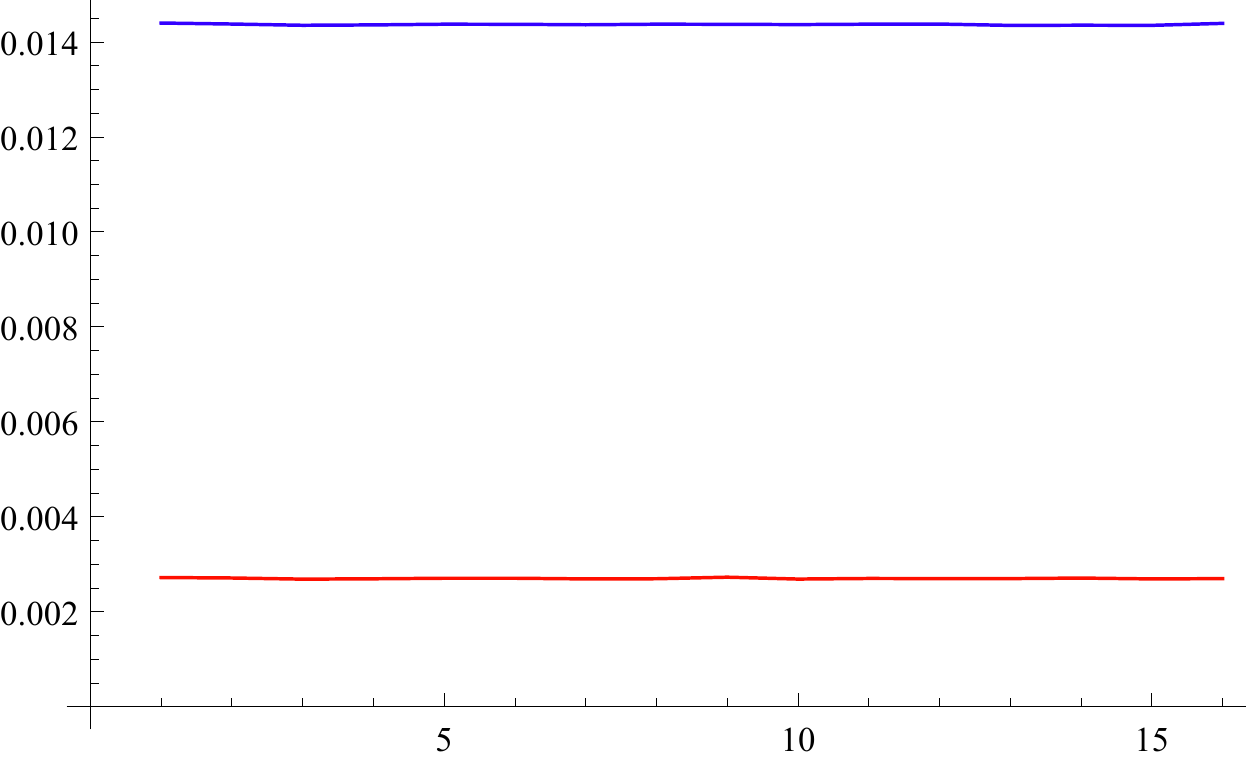}
\caption{Probability of event of interest as a function of $n$ for $k=3$. Blue line corresponds to Laplace distribution; red line - to Gaussian distribution. Sample size changes from 10000 to 25000 with step 1000.}\label{fig4}
\end{figure}

But what model can be used to explain the number of outliers? A set of such models was proposed in \cite{KV}. These are so-called $\nu$-Gaussian distributions playing the role of Gaussian for summation of a random number of random variables. Just the simplest situation is of of Laplace distribution (analog of Gaussian distribution for sums of geometrically distributed number of random variables). The estimator of the probability of interest in this case is given on Figure \ref{fig4}. We see, that this probability is 0.01436. It is close to upper bound of the probabilities from Table 3 in \cite{EK}. However, the use of this model is questionable, and one needs more studies in this field. The attempts to use geometric stable random variables in Finance are known (see, for example, \cite{KP}). Our point of view is that the distributions used for modeling are not supposed to have heavy tails. They have to be oriented of the large number of outliers.

\section{Acknowledgements}
Author is very grateful to Professor J. Huston McCulloch for fruitful discussion and essential corrections.

\end{document}